\documentstyle[aps,prl,epsfig,multicol]{revtex}

\epsfclipon

\begin{document}

\newcommand \be {\begin{equation}}
\newcommand \ee {\end{equation}}
\newcommand \bea {\begin{eqnarray}}
\newcommand \eea {\end{eqnarray}}
\newcommand \nn {\nonumber}

\title{\bf Dynamical ultrametricity in the critical trap model}
\author{E. Bertin, J.-P. Bouchaud}
\address{\it Commissariat \`a l'\'Energie Atomique,
Service de Physique de l'\'Etat Condens\'e\\
91191 Gif-sur-Yvette Cedex, France}

\maketitle

\begin{abstract}
We show that the trap model at its critical temperature presents 
dynamical ultrametricity in the sense of Cugliandolo and Kurchan \cite{CuKu94}.  We use the explicit analytic solution of this model to
discuss several issues that arise in the context of mean-field 
glassy dynamics, such as the scaling form of the correlation
function, and the finite time (or finite forcing) corrections to ultrametricity, that
are found to decay only logarithmically with the associated time scale, as well as the fluctuation dissipation ratio. We also argue that in the multilevel trap model, the short time dynamics is dominated by 
the level which is at its critical temperature, so that dynamical 
ultrametricity should hold in the whole glassy temperature range. We
revisit some experimental data on spin-glasses in light of these
results.\\
\\
\sc{pacs} numbers: 75.10.Nr, 05.20.-y, 02.50.-r
\end{abstract}

\begin{multicols}{2}

\section {Introduction}

Notable theoretical progress in our understanding of the ubiquitous {\it aging}
phenomena in glassy systems was made possible by the recognition that
the discussion of correlation and response functions requires {\it two} times:
the waiting time $t_w$ and the total elapsed time $t_w+t$. This appears very
clearly in the framework of mean-field spin glass models, domain growth (coarsening models),
or more phenomenological trap models \cite{Review}. Although the basic phenomenology of all these 
models are rather similar, the underlying physical picture is completely 
different. For example, aging in domain growth models is associated to the
growth of a coherence length. In mean field or trap models, space is 
absent and aging is related to the structure of the energy landscape, but
here again the intuition is completely different. In mean field models,
the system never reaches the bottom of an energy valley and there are 
no activated processes involved in the dynamics. Rather, the dynamics slows down because saddles with less and less `descending' directions are visited as
the system ages. Conversely, in the trap model, activation is the basic 
ingredient of the model, and aging is associated to the fact that deeper
and deeper valleys are reached as the system evolves. Dynamics in the latter
case in fundamentally {\it intermittent}: either nothing moves, or there is
a jump between two traps. This must be contrasted with mean field dynamics
which is continuous in time. However, mean field and trap dynamics can be shown to correspond to two successive time regimes in a particular class of models \cite{BenArous}.

In spite of these important differences, many predictions are common to the
latter two pictures, such as:
\begin{itemize}
\item A short time singularity of the response function in the aging regime,
which leads to an (aging) low frequency noise.
\item Non trivial violations of the Fluctuation Dissipation Theorem ({\sc fdt}), 
first pointed out 
within mean field models, but that also exists in trap models. 
\item The possibility of rejuvenation and memory, which involves the 
existence of different degrees of freedom with different time scales.
\end{itemize}

Furthermore, a certain class of 
mean field models (that includes the Sherrington-Kirkpatrick model) has been conjectured to possess ultrametric dynamical
properties, that very precisely reflect and encode 
the ultrametric nature of the static solution. This dynamical ultrametricity
is associated to an infinite number of time scales (which diverges
with the age of the system), in the following sense: if $C(t_2,t_1)$ 
is the correlation function between times $t_1 < t_2$, then in the limit
of large times:
\be
C(t_3, t_1) = \min{(C(t_2, t_1), C(t_3, t_2))}, \quad \forall t_2 \in [t_1,t_3].
\ee
This means that either $t_2$ is close enough to $t_3$, and then no 
further dynamics takes place between $t_2$ and $t_3$, or $t_2$ is close 
enough to $t_1$ but then the age of the system hardly changes between 
$t_1$ and $t_2$. This property of the correlation function has been shown to hold for the aging solution of the dynamical (Mode Coupling) equations 
describing the dynamics of `continuous' spin-glasses. This property is furthermore
invariant under reparametrizations of time, where $t \to h(t)$ with
an arbitrary monotonous function $h(t)$. Testing whether or
not dynamical ultrametricity also holds in realistic disordered systems 
is made difficult because this property is only expected 
in the limit of asymptotically large times, and corrections are
expected on finite times. How large are these corrections?

In this paper, we show that exact ultrametricity holds {\it at} the
critical point of the single-level trap model (or random energy model).
We give the explicit form of the correlation function and discuss
finite time (or finite forcing) corrections. 
Note that in this single-level trap model,  
the dynamics is ultrametric although the statics is not. 
The issue of finite time {\sc fdt} plot is also adressed. We discuss
multi-level extensions of the trap model and argue that dynamical 
ultrametricity should be generic at `short times', i.e. at the beginning 
of the aging region. We show thermoremanent magnetization data that
support this idea. The relation with $1/f$ noise, already discussed
in this context \cite{Dean}, is recalled.

\section{The model}

The trap model, introduced in the context of aging in \cite{Bou92,Monthus} and
further developed in \cite{Maass}, 
is one of the simplest soluble model exhibiting a dynamical glass transition. 
In this model, one considers a particle which is trapped in low 
energy states $i$ of depth $E_i > 0$, where the $E_i$ are random
 variables distributed according to $\rho(E)=\frac {1}{T_c} e^{-E/T_c}$. 
The dynamics is chosen to be activated: each particle stays in trap $i$ an
exponential random time, equal on average to $\tau_0 e^{E_i/T}$.
The quantity $\tau_0$ is a microscopic time scale 
which we shall take as the time unit in the following. When the particle leaves 
the trap, it chooses at random a new one among all the others. 
As a consequence, at high temperature ($T>T_c$), the particle spends 
most of its time in the small traps, because the number of these traps 
(the entropic factor) dominates the Boltzmann factor and the system equilibrates. On the contrary, for $T<T_c$, the Boltzmann factor is 
dominant and the particle explores deeper and deeper traps, so that 
the system never equilibrates (in the limit of an infinite number of traps).
In this regime, the dynamics ages: correlation and response functions are
no longer time translation invariant, but depend both on the waiting time
$t_w$ and the total time $t_w+t$.

\section{Correlation function}

Correlation functions are useful tools to characterize the dynamics and to
 compare several models. The simplest (but nevertheless non trivial) correlation 
in the trap model is 
$\Pi(t_w+t, t_w)$ defined as the probability to remain in the same trap during 
the time interval $[t_w, t_w+t]$. As we are considering the infinite 
dimensional (or fully connected) model, this probability is also equal 
to the probability $P(t_w+t, t_w)$ to be in the same trap at $t_w+t$ 
as at $t_w$, since the probability to go back to the same trap vanishes 
in the limit of an infinite number of traps. (This is not true in finite dimension. For 
instance, in one dimension, $\Pi(t_w+t, t_w)$ and $P(t_w+t, t_w)$ scale in a different way with $t_w$).

The correlation function $\Pi(t_w+t, t_w)$ has been calculated in 
the general case in \cite{Dean}. We shall now focus on the critical case 
$T=T_c$, for which we have:
\be
\Pi(t_w+t, t_w) \,=\, \int_1^{t_w} du \frac{1\,-\,\exp{[-(t_w+t-u)]}}{(\log{u})(t_w+t-u)}
\ee
for large $t_w$. We note that this expression simplifies further in the 
limit $t,t_w \to \infty$. 
In particular, the exponential term vanishes since $t_w+t-u > t$. 
Writing $u=t_w (1-v)$, we get
\bea
\Pi(t_w+t, t_w) & \simeq & \int_0^{1-\frac{1}{t_w}} 
\frac {dv}{[\log{t_w} + \log{(1-v)}] [v+\frac{t}{t_w}]}\\
& \simeq & \frac{1}{\log{t_w}} \int_0^{\frac{t_w}{t}} \frac {dz}{z+1} + {\cal O}\left(\frac{1}{(\log{t_w})^2}\right)
\eea
with the new integration variable $z=\frac{t_w}{t}v$. 
This expression is readily integrated to give:
\be \label{coreqn1}
\Pi(t_w+t, t_w) \simeq \frac{\log{(1+\frac{t_w}{t})}}{\log{t_w}}
\ee
This relation shows that $\Pi(t_w+t, t_w)$ 
is not a function of $\frac{t}{t_w}$, at variance with the
results that hold the whole low temperature phase $T < T_c$ \cite{Dean}. 
On the contrary, taking $t \sim a t_w^{\alpha}$, we obtain, 
in the limit $t_w$ going to $+\infty$:
\be
\Pi(t_w+t, t_w) = {\cal C} (\alpha)
\ee
with ${\cal C} (\alpha)$ given by:
\bea
{\cal C} (\alpha) &=& 1-\alpha \quad (\alpha < 1)\\
		&=& 0 \qquad \quad (\alpha \geq 1).
\eea
Note that ${\cal C} (\alpha)$ is a monotonous decreasing function of $\alpha$.
The above result is only true in the limit $t \gg \tau_0$, 
as well as $\log{({t_w}/{\tau_0})} \gg 1$. 
Indeed, all the finite time results reported in this paper should be understood as first order terms in a $1/\log{{t_w}}$ expansion.

An important remark is that the correlation function $\Pi(t_w+t, t_w)$ 
is a function of $\alpha=\log t/\log t_w$ that cannot be written as 
$h(t_w+t)/h(t_w)$. The latter ratio naturally appears (with an 
unknown function $h$) in the aging part of the solution of the dynamical 
equation corresponding to one step Replica Symmetry Breaking ({\sc rsb}) 
mean field spin glass models. However, for full {\sc rsb} models where 
dynamical ultrametricity is indeed expected, the correlation function is 
given by an infinite sum of contributions coming from different times sectors 
\cite{CuKu94,Review}:
\be
C(t_w+t,t_w) = \sum_i {\cal C}_i\left(\frac{h_i(t_w+t)}{h_i(t_w)}\right),\label{sk}
\ee
where $h_i$ are unknown (monotonous) functions defining the $i$th time scale, 
and the ${\cal C}_i$ are monotonously decaying to zero for large arguments. 

A useful (but up to now unjustified theoretically) form for $h_i(t)$, 
that allows one to give some flesh to the above formula, is \cite{Review}:
\be
h_i(u)=\exp\left[\frac{u^{1-\mu_i}}{1-\mu_i}\right], \qquad 0 \leq \mu_i \leq 1.\label{hi}
\ee  
It is easy to see that for this choice of $h_i$, the time scale on which the
ratio $h_i(t_w+t)/h_i(t_w)$ varies significantly is precisely $t_w^{\mu_i}$. 
The choice $\mu=0$ therefore  
corresponds to stationnary dynamics, whereas $\mu=1$ gives full aging. 
Now, take $t=t_w^\alpha$
(with $0 < \alpha < 1$) in Eq. (\ref{sk}) and take the limit $t_w \to \infty$. 
All the sectors such that
$\mu_i < \alpha$ have relaxed to zero, whereas the sectors corresponding to 
$\mu_i > \alpha$ have not decayed at all. Introducing a continuum of different values
of $\mu$, we find that the correlation function is given by:
\be
C(t_w+t_w^\alpha,t_w) = \int_\alpha^1 d\mu \rho(\mu) {\cal C}_\mu(1),
\ee
where $\rho(\mu)$ is the `density' of time sectors of order 
$t_w^\mu$ and ${\cal C}_\mu(1)$ is the
initial value of the correlation function in this sector. From this 
result, one sees that $C(t_w+t,t_w)$ indeed becomes a function of $\alpha=\log t/\log t_w$
in the long time limit. Therefore,
interestingly, the superposition of an infinite number of {\it subaging} 
contributions defined by (\ref{hi}) naturally leads to
a correlation function that depends on $\log t/\log t_w$, for which the dynamical 
ultrametricity  property is explicit. The critical trap behaviour corresponds 
to a uniform contribution of all time sectors, i.e. 
$\rho(\mu) {\cal C}_\mu(1)=1$, $\forall \mu$.

\section{Dynamical ultrametricity}

As recalled in the Introduction, Cugliandolo and Kurchan have
defined dynamical ultrametricity for the correlation function $C$ 
if the following property is true: take three times $t_1 < t_2 < t_3$, 
then:
\be
C(t_3, t_1) = \min{(C(t_2, t_1), C(t_3, t_2))}, \quad \forall t_2 \in [t_1,t_3].
\ee
Let us show now that $\Pi(t_w+t, t_w)$ at the critical temperature 
is ultrametric in the sense defined hereabove. It will be useful to introduce 
the following notations:
\bea
\Pi(t_2, t_1) = C_1\\ \nn
\Pi(t_3, t_2) = C_2\\ \nn
\Pi(t_3, t_1) = C_3 \nn
\eea
From the monoticity of correlation functions, the inequality $C_3 \leq \min{(C_1, C_2)}$ 
holds in general.  We simply have to check that (at least) one of the two correlations 
$C_1$ and $C_2$ is equal to $C_3$. In order to take the infinite time limit, 
we need to specify how $t_2$ and $t_3$ scale with $t_1$. A natural parametrization
is the following:
\bea
t_3 - t_1 &\sim& a\,t_1^{\alpha}\\ \nn
t_2 - t_1 &\sim& b\,t_1^{\beta} \nn
\eea
We can now look at various cases:
\begin{itemize}
\item If $\beta < \alpha$, then one has, for large $t_1$:
\be
t_3 - t_2 \sim a\,t_1^{\alpha}-b\,t_1^{\beta} \sim a\,t_1^{\alpha}
\ee
so that $C_1 = {\cal C} (\beta)$ and $C_2 = C_3 = {\cal C} (\alpha) < {\cal C} (\beta)$.
\item Assuming now $\beta = \alpha$, we get $C_1 = C_3 = {\cal C} (\alpha)$, 
as well as $C_2 \geq C_3$. 
\end{itemize}
As a result, we have shown that the relation $C_3 = \min{(C_1, C_2)}$ always holds, 
which implies that dynamical ultrametricity is satisfied in this model.

The appearance of dynamical ultrametricity can be considered as a signature of 
the existence of many time scales involved in the dynamics. 
This is indeed the case in the present model, even though there is 
no static ultrametricity to account for this hierarchy of time scales. 
This property in fact arises naturally from the exact balance, 
at the critical point, of the Boltzmann weight $e^{E/T}$ and of the entropic factor 
$\rho(E)$: the probability for the particle to have a given energy 
(or equivalently, a given trapping time, in logarithmic scale) is essentially 
uniform on the interval $[0, T_c \log{t_w}]$. 
In other words, dynamical ultrametricity here is a consequence of the critical 
scale invariance.

\section{Finite time analysis}

An interesting analysis was introduced by Cugliandolo and Kurchan in 
the context of the dynamical analysis of the 
Sherrington-Kirkpatrick (SK) model \cite{CuKu94}. 
These authors made the assumption that, in the limit of large times, 
there exists a certain function $f(x, y)$, not necessarily smooth, 
such that $C_3 = f(C_1, C_2)$. We have shown in the previous section that such a 
function indeed exists in the present case and is given by $f(x, y) = \min{(x, y)}$.

A useful representation, proposed in \cite{CuKu94}, is to plot in the $(C_1, C_2)$-plane 
the curves of constant $C_3 = f(C_1, C_2)$, which reduces for our case 
to two straight lines ($C_1 = C_3$ and $C_2 = C_3$) at right angle. 
For finite times, the function $f$ has to include a time scale as third argument, 
so that $C_3 = F(C_1, C_2 ; t_1)$, where $f(x,y)=\lim_{t_1 \to \infty} F(x,y;t_1)$.
The $(C_1, C_2)$-plane representation is then a good way to visualize 
the convergence towards the asymptotic function $f(C_1, C_2)$. 
It has been used with numerical data to test if 
dynamic ultrametricity holds in realistic systems, 
with rather inconclusive results \cite{CuKu94,Berthier}, except in a few cases, like in a recent study of the 4-dimensional Edwards-Anderson model \cite{Stariolo}.

\begin{figure}
\centerline{
\epsfxsize = 8.5cm
\epsfbox{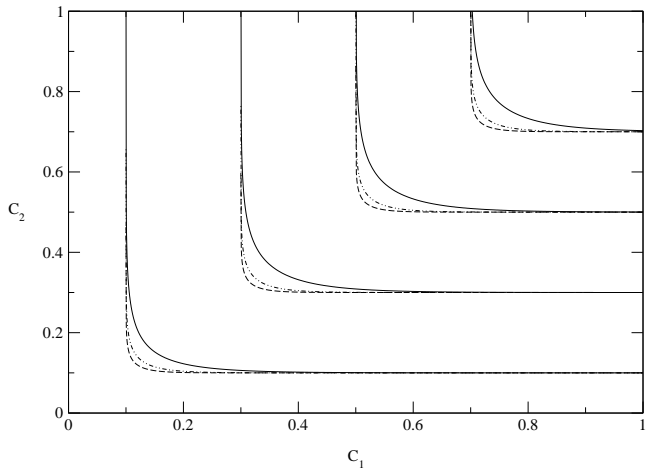}
}
\vskip 0.5 cm
\caption{\sl Plot of constant $C_3 = F(C_1,\,C_2 \,;\,t_1)$ in the $(C_1,\,C_2)$-plane. From left to right, curves correspond to $C_3=0.1, 0.3, 0.5, 0.7$. Each set of three curves shows the convergence with $t_1$ towards the asymptotic function $f(C_1, C_2) = \min{(C_1, C_2)}$: $t_1=10^5$ (full line), $10^{10}$ (dot-dashed line), $10^{15}$ (dashed line).}
\label{corfig1}
\end{figure}

Let us apply this procedure to the critical trap model. In order to deal 
with finite time expressions, we shall come back to Eqn. \ref{coreqn1}, restated as follows:
\be
C(t_j, t_i) = \frac{\log{(1+\frac{t_i}{t_j-t_i})}}{\log{t_i}}
\ee
Inverting these relations so as to express $t_2$ and $t_3$ as functions of 
$t_1$, $C_1$ and $C_3$, we can write an explicit expression for $F(C_1, C_2; t_1)$:
\bea \nn
C_3 &=& F(C_1, C_2; t_1)\\
  &=& -\frac{1}{\log{t_1}} \log{\left(t_1^{-C_1}+t_1^{-C_2}(1-t_1^{-C_1})^{1+C_2}\right)}
\eea
In order to plot the constant $C_3$ curves, 
it will be useful to express also $C_2$ as a certain function $G(C_1, C_3; t_1)$:
\be
C_2 = G(C_1, C_3; t_1) = \frac{\log({1-t_1^{-C_1}})-\log(t_1^{-C_3}-t_1^{-C_1})}
{\log{t_1}-\log(1-t_1^{-C_1})}
\ee
The resulting plots in the $(C_1, C_2)$-plane are displayed in 
Fig. \ref{corfig1}, for $C_3 = 0.1, 0.3, 0.5, 0.7$ and for 
three different times ($t_1 = 10^5$, $10^{10}$ and $10^{15}$). Note that the
convergence is very slow close to the infinite time singularity.

\begin{figure}
\centerline{
\epsfxsize = 8.5cm
\epsfbox{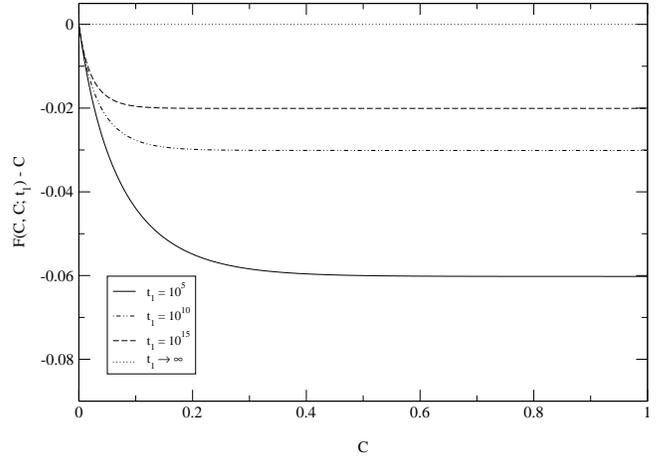}}
\vskip 0.5 cm
\caption{\sl Plot of $F(C, C; t_1)-C$ versus $C$ for the same values of $t_1$ as  in Fig. \ref{corfig1}. The dotted curve, corresponding to the infinite time limit $f(C,C)=C$ is added for comparison. The departure from $f(C, C)=C$ is almost independent of $C$, except for $C$ close to $0$.}
\label{corfig2}
\end{figure}

Cugliandolo and Kurchan also introduced in \cite{CuKu94} the notion of 
correlation time scales through the representation of $f(C, C)$ versus $C$. 
As already mentionned before, $f(C, C) \leq C$ in general. There may 
exist some special `fixed' points $C^*$ such that $f(C^*, C^*) = C^*$. 
Each of these fixed points has been shown to be associated with a correlation time scale. 
If dynamical ultrametricity holds in a particular time sector, all $C$'s belonging 
to a certain interval $[C', C'']$ are fixed points. 
In our case, ultrametricity holds over the full correlation interval $[0, 1]$. 
But in our model, we can go beyond the infinite time analysis 
and quantify the convergence of 
$F(C, C; t_1)$ with $t_1$ towards the asymptotic function $f(C, C) = C$. We find:
\be
F(C, C; t_1) = C\,-\,\frac{\log{(1+(1-t_1^{-C})^{1+C})}}{\log{t_1}}
\ee
For $C>0$, this expression simplifies further at large times:
\be \label{conv}
F(C, C; t_1) \simeq C\,-\frac{\log{2}}{\log{t_1}}
\ee
Interestingly, the leading correction does not depend on $C$, and it is valid only if $C$ is not too close to $0$. Fig. \ref{corfig2} displays the plots of $F(C, C; t_1)-C$, for the same values of $t_1$ as in Fig. \ref{corfig1}.

This last result may also be interpreted in the framework of Fig. \ref{corfig1}. 
For a given value of $C_3$, the point $C(t_1)$ defined by the relation 
$F(C(t_1), C(t_1); t_1)=C_3$ converges to the infinite time right angle singularity $C_1=C_2=C_3$ as (see Eqn. \ref{conv}):
\be
C(t_1) \simeq C_3\,+\,\frac{\log{2}}{\log{t_1}}
\ee
This logarithmic correction may explain why it seems so difficult to observe the 
convergence towards the ultrametric relation in experimental or numerical data, 
where only a few decades (usually between four and six) are available.

\section{The effect of `shear'}

The effect of an external `shear' (or power injection) on aging 
was investigated in the context of mean field models in \cite{Berthier},
and in the context of the trap model in \cite{SGR}. In both models, aging
is interrupted by the shear beyond a time scale $\tau_r$ which diverges as the shear rate $\dot \gamma$ tends to zero. In the model considered in \cite{SGR}, this time scale is given by:
\be
\tau_r \simeq \frac{1}{\dot \gamma} \left(\log \frac{1}{\dot \gamma}\right)^{\frac{1}{2}}.
\ee
For $\tau \ll \tau_r$, the (power-law) distribution of trapping times is in a 
first approximation unaffected by the shear, whereas for longer times, the distribution decays exponentially. In the limit where the waiting time $t_w$ is much larger than $\tau_r$, the dynamics of the model becomes stationary and one finds, for $t \ll \tau_r$, the same result as above with $t_w$ replaced by
$\tau_r$:
\be
C(t+t_w,t_w)={\cal C}\left(\frac{\log t}{\log \tau_r}\right) \simeq 1 - \frac{\log{t}}{\log{\tau_r}}.
\ee 
As discussed in \cite{Berthier,Stariolo}, dynamical ultrametricity in this context 
manifests itself by the appearance of an infinity of time scales in the
limit $\tau_r \to \infty$ (i.e. $\dot \gamma \to 0$): the time needed for
the correlation to decay to a certain value $c$ diverges as $\tau_r^{1-c}$
(see \cite{Kurchan} for a further discussion).

\section{The fluctuation-dissipation ratio}

\begin{figure}
\centerline{
\epsfxsize = 8.5cm
\epsfbox{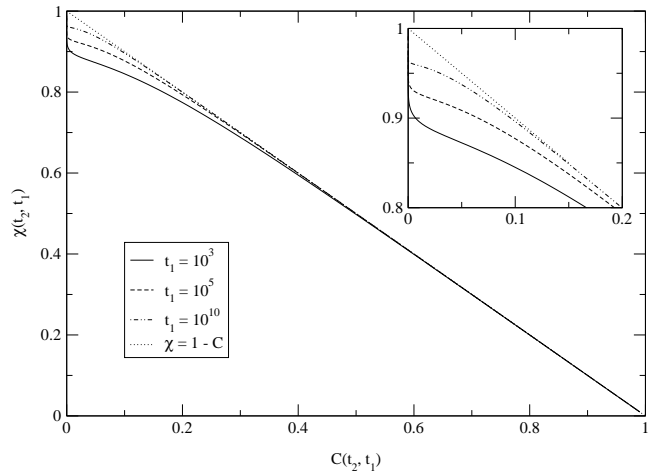}}
\vskip 0.5 cm
\caption{\sl Plot of integrated response $\chi (t_2, t_1)$ versus $C(t_2, t_1)$ parametrized by $t_2$, for $t_1 = 10^{3}$, $10^{5}$ and $10^{10}$ from bottom to top. The local slope is equal to $-X(t_2, t_1)/T_c$, and it converges very slowly towards the asymptotic value $-1/T_c$ for small $C$ (see inset). $T_c$ is chosen as the temperature unit.}
\label{fdt}
\end{figure}

It is interesting to study the fluctuation dissipation ratio $X$ in the trap
model at the critical point. This ratio is defined as:
\be
X(t_2,t_1)= \frac{TR(t_2,t_1)}{\partial C(t_2,t_1)/\partial t_1},
\ee
where $R(t_2,t_1)$ is the response of the system at time $t_2>t_1$ to a small bias field 
applied at time $t_1$ (see \cite{Dean,Fielding} for details). In the trap
model, where the field only changes the trapping time of the starting site,
one finds that the following relation holds in general:
\be
T\, R(t_2,t_1)=-\frac{\partial C(t_2,t_1)}{\partial t_2}.
\ee
Using this result, one sees that in the `liquid' phase $T>T_c$ where all 
two time functions only depend on time differences, $X \equiv 1$: the 
usual Fluctuation-Dissipation Theorem ({\sc fdt}) holds. In the glass phase, on 
the other hand, one finds that $X(t_2,t_1)=t_1/t_2$: the value of
$X$ is non trivial in the whole scaling regime $t_2-t_1 \sim t_1$. Right at the 
critical point $T=T_c$, one can express $X$ as:
\be
X(t_2,t_1) = \frac{(t_1^C-1)^2}{t_1^C(t_1^C+1-C)} \qquad C \equiv C(t_2,t_1) > 0.
\ee
So for any fixed $C>0$, $X$ tends to $1$ in the asymptotic
limit $t_1 \to \infty$. We show
in Fig. \ref{fdt} the now famous plot of the integrated response versus $C$, which should yield a straight line of slope $-1/T_c$ when the {\sc fdt} holds. Again,
one sees a very slow convergence towards the asymptotic result for small values of $C$.

\section{The multilevel trap model and discussion}

The simple trap model described above can be considered as a one-step 
{\sc rsb} model. 
In order to generalize the model to a full {\sc rsb} one, it has been proposed in 
\cite{Dean} (and further studied in \cite{Sasaki}) to follow Parisi's procedure 
for the static solution of the SK model. Roughly speaking, it means that each trap is 
recursively subdivided into a new series of traps, in a hierarchical manner. 
Each level $k$ of traps is characterized a certain overlap between states $q_k$ and by an exponential probability distribution 
of the energy barriers, with a critical temperature $T_c^k$ depending on the level index 
$k$. The critical temperatures are related to Parisi's function $x_k=x(q_k)$ 
as $T_c^k=T/x_k$, and satisfy the relations $T_c^k < T_c^{k-1}$.  At any temperatures,
the levels of the tree corresponding to $q > q_{EA}(T)$, where $q_{EA}$ is the 
Edwards-Anderson parameter, are such that $x_k > 1$, so that these levels are 
equilibrated (i.e. $T_c^k < T$).

In the single level trap model, the correlation function 
$\Pi(t_w+t, t_w)$ in the aging phase $T < T_c$ behaves at short times as:
\be
\Pi(t_w+t, t_w) \simeq 1-\frac{\sin{\pi x}}{\pi (1-x)}\,\left(\frac{t}{t_w}\right)^{1-x} \quad 1 \ll t \ll t_w
\ee
with $x = \frac{T}{T_c}$.
In the multi-level model with a finite number $M$ of levels, 
the total correlation function is defined as:
\bea\label{corrtree} \nn
C(t_w+t, t_w)&=&\sum_{k=0}^M q_k \, [\Pi_k(t_w+t, t_w)-\Pi_{k+1}(t_w+t, t_w)]
\\ &=&q_0 + \sum_{k=1}^M (q_k-q_{k-1}) \, \Pi_k(t_w+t, t_w)
\eea
$q_k$ being the $k^{th}$ level overlap, and $\Pi_{k}(t_w+t, t_w)$
is the probability that the process has never jumped beyond the $k^{\rm th}$ layer of the tree between $t_w$ and $t_w+t$ with the convention that $\Pi_0(t_w+t, t_w) = 1$ and $\Pi_{M+1}(t_w+t, t_w) = 0$ (see \cite{Dean} for details). From this definition, we see that $C(t_w+t, t_w)$ is dominated at short times by the levels $k$ with $x_k$ close to $1$, for
which the short time singularity is strongest. (We assume that $T < T_c^0$, i.e. that at least one level is aging). Therefore, we expect to 
observe the dynamical ultrametricity associated to the level $k^*$ for 
which $x_{k^*}=1$ in the `short' time regime $\log t/\log t_w < 1$, before the
$t/t_w$ regime associated to the levels $k < k^*$ sets in. (Note that 
if $C(t_w+t, t_w)$ is a function of $t/t_w$, then the function $f(x,y)$ 
defined above cannot be equal to $\min(x,y)$).

Interestingly, `short time' dynamical ultrametricity exists for the
hierarchical tree model in the whole low temperature phase, but has no 
relation with the static ultrametricity built in the tree structure which
encodes Parisi's {\sc rsb} solution. Thus the origin
of dynamical ultrametricity in the generalized trap model is again very 
different from the dynamical ultrametricity found in mean field spin glass
models, which in the latter case is deeply related to the Parisi function
$x(q)$ which encodes the structure of the tree. As mentioned in the introduction, the physical interpretation of aging in the two pictures
are radically different, although some of the phenomenology is very similar.

\begin{figure}
\centerline{
\epsfxsize = 8.5cm
\epsfbox{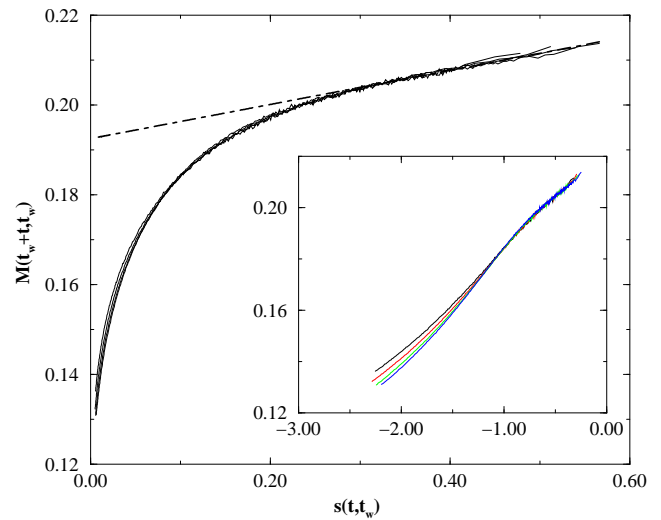}}
\vskip 0.5 cm
\caption{\sl Plot of {\sc trm} data in a AgMn spin-glass at $T/T_g=0.75$, 
for $t_w=300$, $1000$, $3000$ and $10000$ seconds. The horizontal axis is 
the variable $s(t,t_w)$ defined in the text. The rescaling is very good for 
 $s(t,t_w) > 0.3$ approximately (or $\log t/\log t_w < 0.7$), but 
becomes inadequate for longer times $t$, as more clearly seen in the inset 
where $\log s$ is used. The dashed line is an affine fit of the short time
 part of the data.}
\label{agmn}
\end{figure}

Aging experiments is spin-glasses have been interpreted within the framework
of the multi-level trap model in \cite{Dean}. The need for several levels 
comes from rejuvenation and memory in temperature shift experiments, but also from the detailed shape of the thermo-remanent magnetization ({\sc trm}) relaxation at a given temperature,
which shows that the short time and long time singularities are described
by different exponents $x_k$ \cite{Sitges}. It is natural to interpret the different 
levels of the hierarchy in terms of length scales, as is actually suggested
by the {\sc rsb} solution of the pinned manifold problem \cite{BBM}. Therefore,
one can expect that for any temperature within the spin-glass phase, there
will be a particular `critical' length scale $\ell$ for which $x=x(\ell)=1$,
that will contribute to the correlation and response functions as 
a function of  $\log t/\log t_w$. We have reanalyzed some {\sc trm} data in
the spirit of the present discussion. We show in Fig. (\ref{agmn}) the 
decay of the {\sc trm} in AgMn at $T=0.75 T_g$ \cite{AgMn}, plotted as a function of $s(t,t_w)=\log (1+t_w/t)/
\log t_w$, as suggested by Eq.~(\ref{coreqn1}). This figure shows that 
the rescaling is very good at short times, but is violated for large $t/t_w$
(corresponding to small $s(t,t_w)$). From Eq.~(\ref{corrtree}), 
we indeed expect to observe the sum of a contribution ${\cal M}(k < k^*)$ from the levels $k < k^*$ 
(that only depends on $t/t_w$) and a contribution from $k \simeq k^*$, proportional
to $s(t,t_w)$. When $\alpha=\log t/\log t_w < 1$, the first contribution does not
vary much, since $(t/t_w)^{1-x_k} = t_w^{(\alpha-1)(1-x_k)}$ is very small for large $t_w$, whereas the second contribution is a function of $\alpha$. 
This suggests that in the short time regime one should observe:
\be
M(t_w+t,t_w)= {\cal M}(k < k^*) + m_1 s(t,t_w),
\ee
where $m_1$ is a certain prefactor. The quantity $\varphi=m_1/({\cal M}(k < k^*) + m_1)$
measures the relative contribution of the levels $ < k^*$ and around $k^*$
in the total decay of the signal. The dashed line shown in Fig. \ref{agmn}
is a linear fit of the initial decay, as a function of $s$, from which
one extracts (in this particular example) $\varphi \approx 0.2$.

In all the above formulas, time is implicitely measured in units of a microscopic time $\tau_0$. The value of $\tau_0$ is not necessarily an
individual flip time $\sim 10^{-12}$ sec., since collective dynamics may exist 
in the vicinity of the transition point (see \cite{Dupuis,JPB} for a detailed
discussion). In Fig. \ref{agmn}, we have chosen $\tau_0=10^{-5}$ sec. to 
achieve the best rescaling. This value is very close to the one extracted 
from the analysis of \cite{Dupuis}.

The conclusion is that the {\sc trm} data is indeed compatible with a 
$\log t/\log t_w$ behaviour for short times. Note that this behaviour is
tantamount to a logarithmic dependence of the a.c. susceptibility, or
else to $1/f$ noise. We have insisted in previous papers \cite{Dean,Sasaki} on the fact that
the existence of levels in the vicinity of $x=1$ generically leads to 
$1/f$ noise for long times and low frequencies, because the contribution
of other levels (both above and below $k^*$) become negligible in that regime.
We show here that the degrees of freedom contributing to $1/f$ noise also
give rise to dynamical ultrametricity.

\section*{Aknowledgments} We thank L. Berthier and E. Vincent for fruitful discussions.

\end{multicols}


\begin{thebibliography}{99}


\bibitem{CuKu94}
L.~F. Cugliandolo, J. Kurchan, J. Phys. A {\bf 27}, 5749 (1994).

\bibitem{Review}
J.-P. Bouchaud, L. Cugliandolo, J. Kurchan, M. M\'ezard, in {\it Spin-glasses and Random Fields}, edited by A.~P. Young (World Scientific, Singapore, 1998), and references therein.

\bibitem{BenArous}
G. Ben Arous, A. Bovier, V. Gayrard, preprint cond-mat/0110223.

\bibitem{Bou92}
J.-P. Bouchaud, J. Phys. I, {\bf 2}, 1705 (1992).

\bibitem{Monthus}
C. Monthus, J.-P. Bouchaud, J. Phys. A {\bf 29}, 3847 (1996).

\bibitem{Maass}
B. Rinn, P. Maass, J.-P. Bouchaud, Phys. Rev. B {\bf 64}, 104417 (2001).

\bibitem{Dean}
J.-P. Bouchaud, D.S. Dean, J. Phys. I France {\bf 5}, 265 (1995).

\bibitem{Berthier}
L. Berthier, J.~L. Barrat, J. Kurchan, Phys. Rev. E {\bf 63}, 016105 (2000).

\bibitem{Stariolo}
D.~A. Stariolo, Europhys. Lett. {\bf 55}, 726 (2001).

\bibitem{SGR} P. Sollich, F. Lequeux, P. Hebraud, M. Cates;
Phys. Rev. Lett. {\bf 70} 2020 (1997), P. Sollich, Phys. Rev. E {\bf 58}, 738 
(1998), S. M. Fielding, P. Sollich, M. Cates, J Rheology {\bf 44} 323 (2000).

\bibitem{Kurchan} J. Kurchan, preprint cond-mat/0110628.

\bibitem{Fielding} P. Sollich, S. Fielding, P. Mayer, preprint cond-mat/0111241.

\bibitem{Sasaki} M. Sasaki and K. Nemoto, J. Proc. Soc. Jpn. {\bf 69} 2283 (2000)

\bibitem{Sitges} E. Vincent, J. Hammann, M. Ocio, J.-P. Bouchaud and L.F. 
Cugliandolo:
in {\it Proceeding of the Sitges Conference on Glassy Systems},
Ed.: E. Rubi (Springer, Berlin, 1996).

\bibitem{BBM}
L. Balents, J.-P. Bouchaud, M. M\'ezard, J. Phys. I {\bf 6}, 1007 (1996).

\bibitem{AgMn}  M. Alba, M. Ocio, J. Hammann, Europhys. Lett. {\bf 2}, 45 (1986), J. Phys.
Lett. {\bf 46} L-1101 (1985), M. Alba, J. Hammann, M. Ocio, Ph. Refregier, J. Appl. Phys. 
61, 3683 (1987), E. Vincent, J. Hammann, M. Ocio, p. 207 in
"Recent Progress in Random Magnets", D.H. Ryan Editor, (World
Scientific Pub. Co. Pte. Ltd, Singapore 1992).

\bibitem{Dupuis} V. Dupuis, E. Vincent, J.-P. Bouchaud, J. Hammann, A. Ito, H. Aruga Katori, Phys. Rev. {\bf B 64} 174204
 (2001).

\bibitem{JPB} J.-P. Bouchaud, V. Dupuis, E. Vincent, J. Hammann, to appear in 
Phys. Rev. B.


\end{thebibliography}
\end{document}